\documentclass[pra,twocolumn,aps,amssymb,footinbib]{revtex4-1}
\usepackage{graphicx}
\usepackage{amsmath,amssymb}

\newcommand{\ket}[1]{\ensuremath{\left| #1 \right\rangle}}
\newcommand{\bra}[1]{\ensuremath{\left\langle #1 \right|}}
\newcommand{\ketbra}[2]{\ensuremath{\left| #1 \right\rangle\left\langle #2 \right|}}

\begin{document}

\title{Realizing Fractional Chern Insulators with Dipolar Spins}

\author{N. Y. Yao$^{1\dagger*}$, A. V. Gorshkov$^{2\dagger}$, C. R. Laumann$^{1,3\dagger}$, A. M. L\"{a}uchli$^{4}$, J. Ye$^{5}$, M. D. Lukin$^{1}$}

\affiliation{$^{1}$Physics Department, Harvard University, Cambridge, MA 02138, U.S.A.}
\affiliation{$^{2}$Institute for Quantum Information and Matter, California Institute of Technology, Pasadena, CA 91125, U.S.A.}
\affiliation{$^{3}$ITAMP, Harvard-Smithsonian Center for Astrophysics, Cambridge, MA 02138, U.S.A.}
\affiliation{$^{4}$Institute for Theoretical Physics, University of Innsbruck, A-6020 Innsbruck, Austria}
\affiliation{$^{5}$JILA, National Institute of Standards and Technology and University of Colorado, Department of Physics, University of Colorado, Boulder, Colorado 80309, USA}
\affiliation{$^{\dagger}$These authors contributed equally to this work}
\affiliation{$^{*}$e-mail: nyao@fas.harvard.edu}

\begin{abstract}
Strongly correlated quantum systems can exhibit exotic 
behavior controlled by 
topology. 
We predict that the $\nu=1/2$ fractional Chern insulator arises naturally in a two-dimensional array of driven, dipolar-interacting spins.  
As a specific implementation, we analyze how to prepare and detect synthetic gauge potentials for  the rotational excitations of ultra-cold polar molecules trapped in a deep optical lattice. 
While the orbital motion of the molecules is pinned, at finite densities, the rotational excitations form a fractional Chern insulator.
We present a detailed experimental blueprint for $^{40}$K$^{87}$Rb, and demonstrate that the energetics are consistent with near-term capabilities. 
Prospects for the realization of such phases in solid-state dipolar systems are discussed as are their possible applications.
\end{abstract}
\maketitle


The quest to realize novel forms of  topological quantum matter has recently been galvanized by the theoretical prediction and subsequent experimental observation of topological insulators \cite{Bernevig06, Koenig07}. 
Such materials harbor an insulating bulk, but owing to  nontrivial bulk topology, they are also characterized by robust conducting surface states. 
Recent theoretical work has shown that combining single particle topological bands with  strong interactions  can yield so-called fractional Chern insulating (FCI) phases \cite{Parameswaran11, Regnault11, Sheng11,   Liu12,Wang11,McGreevy12}. 
Particles injected into these exotic states of matter fractionalize into multiple independently propagating pieces, each of which carries a fraction of the original particle's quantum numbers. 
Unlike traditional bosons or fermions, these anyonic excitations accumulate a nontrivial phase under exchange.  

While similar effects underpin the fractional quantum Hall effect observed in continuum two dimensional electron gases \cite{Stormer99}, fractional Chern insulators, by contrast,  are lattice dominated.  They have an extremely high density of correlated particles whose collective  excitations  can transform non-trivially under lattice symmetries \cite{McGreevy12,Wen02}. 
In this paper, we predict the existence of a FCI state  in dipolar interacting spin systems (see Fig.~1). This state exhibits fractionalization of the underlying spins into quasiparticle pairs with semionic statistics \cite{Laughlin87}. The predicted FCI state may also be viewed as a gapped chiral spin liquid \cite{Laughlin87,Nielsen12}, a state  which has never been observed in nature.  Such a state cannot be realized in conventional electron gases. 


Several  recent  studies have conjectured the existence of fractionalized topological phases in idealized lattice models that require sensitively tuned long-range hopping and interactions \cite{Liu12,Sheng11,Wang11,Sun11,Tang11, Neupert11}.  
Broadly speaking, two single-particle microscopic ingredients are required, both of which find close analogy in the physics of the electronic Hall effect.
First, just as electrons in a Landau level have no dispersion, the dispersion of the lattice band-structure must be quenched relative to the energy scale of interactions.
This enables interactions between particles to dominate over the kinetics of their environment \cite{Sun11,Tang11, Neupert11}. 
Second, the flat lattice band should possess a non-trivial Chern number, reflecting the underlying Berry phase accumulated by a particle moving in the band-structure. 
In the context of electronic systems, this corresponds to the well-known quantization of the Hall conductance $\sigma_{xy}$.
To this end, the particles in a Chern insulator need also be coupled to a background gauge potential that breaks time-reversal.
Finally, to observe a \emph{fractionalized} insulating state, one must partially fill the topological flat band-structure with interacting particles; since the FCI state generally competes with superfluid and crystalline orders, the resulting phase diagram naturally exhibits both conventional and topological phases (Fig.~1). 
Up to now, it has been unclear whether such exotic fractional Chern insulating phases can be realized in any real-world physical system. 

\begin{figure*}
\begin{center}
\includegraphics[width=0.6\textwidth]{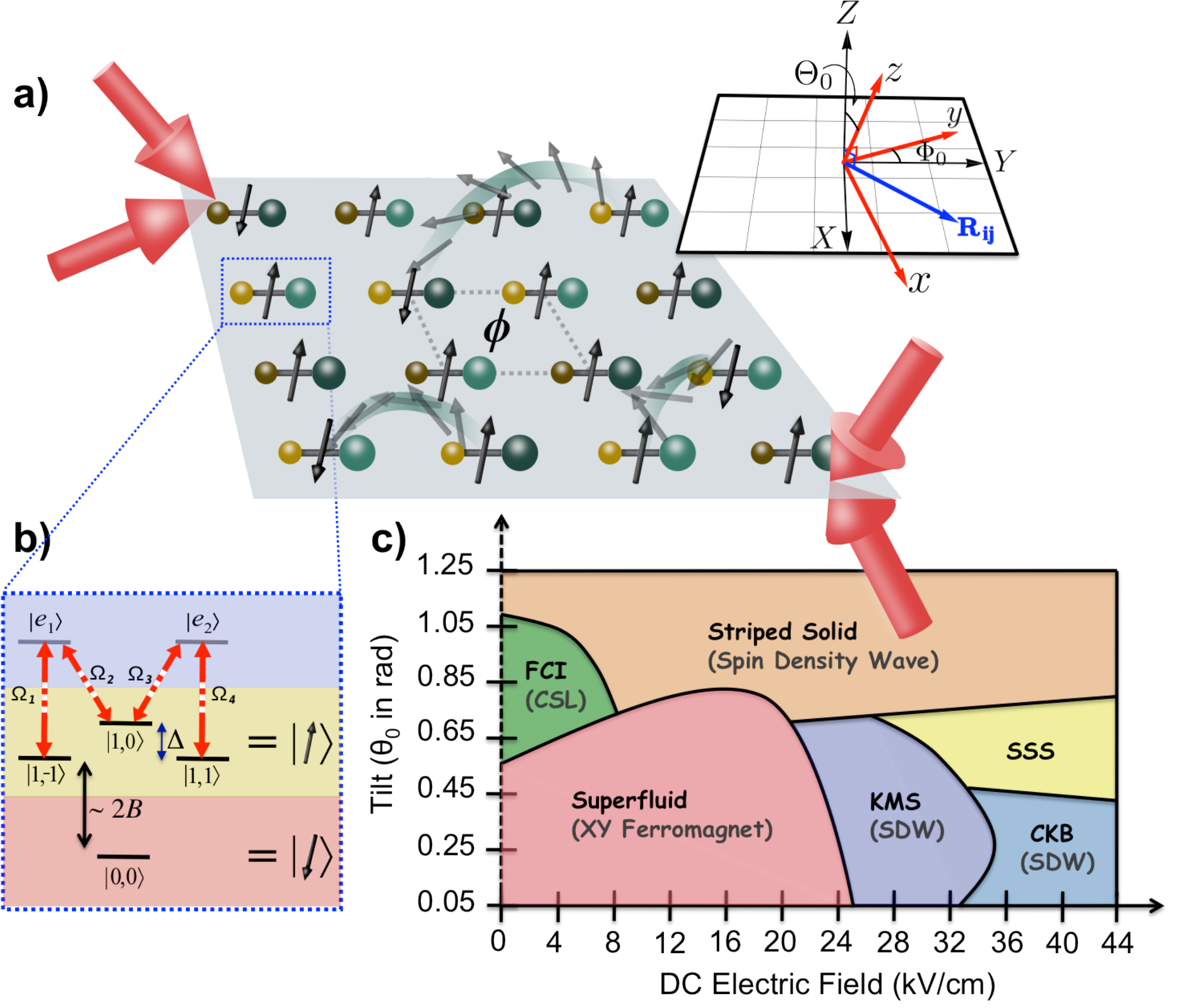}
\end{center}
\caption{\textbf{Realization of a fractional Chern insulator.} {\bf (A)} Schematic representation of the two-dimensional array of polar molecules dressed by optical beams (red arrows). Each polar molecules is characterized as an effective pseudo-spin-flip, which can hop and interact mediated by the long-range dipolar interaction; $\phi$ represents the Aharanov-Bohm phase which the spin-flip acquires as it traverse a plaquette.  (inset) Molecules occupy the $\{X,Y\}$ plane and the rotational quantization axis is set by an applied electric field along the $\hat{z}$ direction. $\Theta_0$ and $\Phi_0$ define the $\{x,y,z\}$ axes with respect to the lattice coordinates $\{X,Y,Z\}$.
{\bf (B)} We consider the $J=0,1$ manifold of each molecule with the $|0,0\rangle$ state representing spin-down. The spin-up state is created via optical Raman dressing in the $M$-configuration. The optical radiation admits a single  dark eigenstate, which is a linear combination of the three states in the $J=1$ manifold. {\bf (C)} Phase diagram for ${}^{40}$K${}^{87}$Rb molecules at half-filling with a total of $N_s=24$ sites as a function of electric field strength and tilt $\Theta_0$. Each phase finds a direct analogy in the language of frustrated magnetism and the equivalent nomenclature is given below. The dotted line at $|E|=0$ signifies the fact that a minimal electric field is always required to split the degeneracy within the $J=1$ manifold.  }
\label{fig:mscheme}
\end{figure*}


We consider a two-dimensional array of  tilted, driven, generalized spins interacting exclusively through their intrinsic dipolar interaction, as depicted in Fig.~1A. 
This interaction mediates the long-range hopping of spin-flip excitations, and  both  of the single-particle ingredients sketched above arise naturally in this context. 
The quenching of the spin-flip band-structure owes to the anisotropy of the dipole-dipole interaction, which yields interference between different hopping directions \cite{Yao12}. 
The production of a synthetic  background gauge potential is accomplished via spatially varying electromagnetic radiation \cite{Yao12, Spielman11}. 
Together, the dipolar anisotropy and radiation induce orientation-dependent Aharonov-Bohm phases that ultimately generate topologically nontrivial  flat bands \cite{Yao12}.

Our approach is applicable to generic pseudospin degrees of freedom, including singlet-triplet states associated with solid-state electronic spin dimers, hyperfine coupled electronic and nuclear spins and cold atomic systems \cite{Giamarchi08,Childress06,ni08}.  To be specific, here, we focus on an implementation using ultra-cold polar molecules trapped  in a deep two-dimensional optical lattice. Such an implementation has many advantages, including local spatial addressing, stable long-lived spins, high phase-space density and intrinsic dipolar interactions \cite{Chotia12,ni08}.
The molecules are subject to a static electric field $\vec{E}$ tilted with respect to the lattice plane (Fig.~1A). 
We assume that the molecular motion is pinned, and hence, restrict our attention to an effective rotational degree of freedom on each site. In particular, we focus on the four lowest rotational levels: $|0,0\rangle$, the rovibrational ground state and  the three states within the $J=1$ manifold ($|1,-1\rangle$, $|1,0\rangle$, $|1,1\rangle$), where
 $J$ characterizes the rotational angular momentum of the molecules. Here,  the quantization axis, $\hat{z}$,  lies along the applied electric field and $\ket{J,m}$ denotes the state adiabatically connected (via $\vec{E}$) to the rotational eigenstates \cite{Gorshkov11}.
Each molecule is driven by optical radiation, which couples the three $J=1$ states to  a pair of molecular excited states $|e_1\rangle$ and $|e_2\rangle$, in the so-called $M$-scheme (Fig.~\ref{fig:mscheme}B). 
The Hamiltonian for each molecule with the laser on has the form
\begin{align}
\label{eq1}
H_r &= \hbar  [ \hspace{1mm}|e_1\rangle (\Omega_1 \langle 1,-1| + \Omega_2 \langle 1,0|)  \nonumber \\
 &+ |e_2\rangle (\Omega_3 \langle 1,0| + \Omega_4\langle 1,1|) + \text{h.c.} \hspace{1mm}  ]
\end{align} 
in the rotating frame, where $\Omega_i$ are Rabi frequencies serving as the control parameters, and h.c.~represents the hermitian conjugate terms. 
The above Hamiltonian admits a unique  ``dark'' eigenstate, $\ket{\uparrow} = \frac{1}{\tilde{\Omega}}( \Omega_2\Omega_4 |1,-1\rangle -\Omega_1\Omega_4 |1,0\rangle + \Omega_1\Omega_3 |1,1\rangle )$, which is  decoupled both from the excited states and from the radiation field ($\tilde{\Omega}$ is a normalization).  Together with the rovibrational ground state, which we label as $\ket{\downarrow}$, this forms an effective two-state spin degree of freedom on each site \cite{Yao12, Gorshkov11, Micheli06,Manmana12}.




Individual molecules interact with one another via electric dipole-dipole interactions, 
\begin{equation}
H_{dd} = \frac{1}{2} \sum_{i\neq j}  \frac{\kappa}{R_{ij}^3}   \left [  {\bf d}_i  \cdot {\bf d}_j - 3({\bf d}_i \cdot {\bf \hat{R}}_{ij})({\bf d}_j \cdot {\bf \hat{R}}_{ij}) \right ],
\end{equation} 
where $\kappa = 1/(4\pi\epsilon_0)$ and ${\bf R}_{ij}$ connects molecules $i$ and $j$ (with dipole moment operators ${\bf d}_i$ and ${\bf d}_j$). 
We let $d$ be the permanent molecular dipole moment and $R_0$ be the nearest-neighbor lattice spacing; by ensuring that the characteristic dipolar interaction strength, $\kappa d^2/R_0^3$, is much weaker than the optical dressing, $\Omega_i$, all molecules remain within the Hilbert space spanned by $\{  \ket{\uparrow} , \ket{\downarrow}\}$. 
Moreover, this interaction  is also  much weaker than the bare rotational splitting $2B$ (Fig.~1B) and thus cannot cause transitions that change the total number of $\ket{\uparrow}$ excitations. 
This effective conservation law suggests the utility of recasting the system in terms of hardcore bosonic operators, $a_i^{\dagger} = \ketbra{\uparrow}{\downarrow}_i$, which create spin-flip ``particles''. 
Mediated by the dipolar interaction, these molecular spin-flips hop from site $j$ to site $i$ with amplitude $t_{ij} =  -\bra{\uparrow_i \downarrow_j} H_{dd} \ket{ \downarrow_i \uparrow_j}$. As each hardcore boson harbors an electric-field induced  dipole moment, there also exist long-range density-density interactions of strength  $V_{ij} = \bra{\uparrow_i\uparrow_j} H_{dd} \ket{ \uparrow_i\uparrow_j} +  \bra{\downarrow_i\downarrow_j} H_{dd} \ket{ \downarrow_i\downarrow_j } -   \bra{\uparrow_i\downarrow_j} H_{dd} \ket{ \uparrow_i\downarrow_j} -  \bra{\downarrow_i\uparrow_j } H_{dd}\ket{ \downarrow_i\uparrow_j}$. In combination, this yields a two-dimensional model of   hardcore lattice bosons,  
\begin{equation}
	\label{eq:ham_hcboson}
H_B =  -\sum_{ij} t_{ij} a_i^{\dagger} a_j + \frac{1}{2}\sum_{i \neq j} V_{ij} n_i n_j,
\end{equation}
whose total number, $N = \sum_i a_i^{\dagger} a_i$, is conserved \cite{Yao12}. Variations in the dipolar-induced on-site potential, $t_{ii}$, can be regulated via tensor shifts from the optical lattice (see supporting information for details).



\begin{figure}
\begin{center}
\includegraphics[width=3.4in]{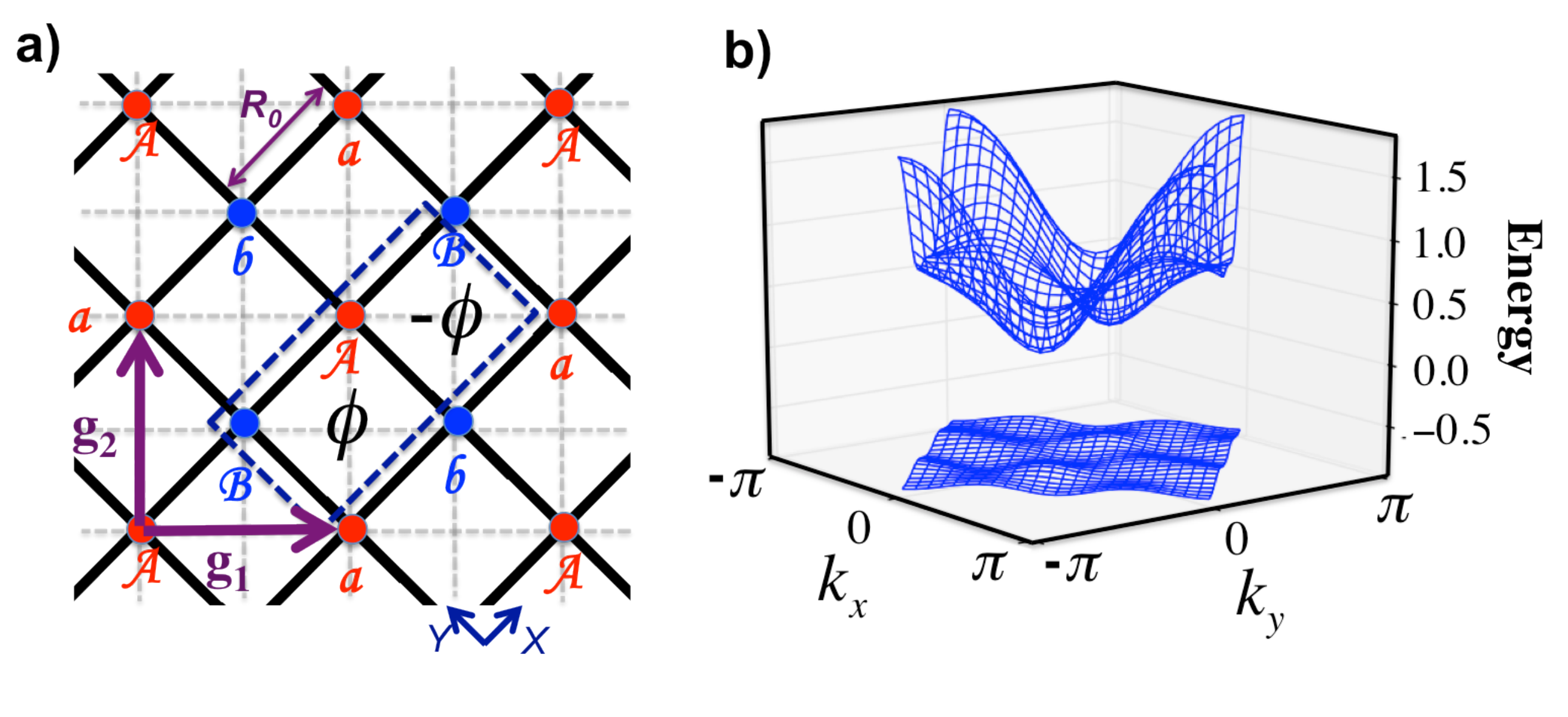}
\end{center}
\caption{\textbf{Topological Flat Bands.} {\bf (A)} Schematic representation of the 2D dipolar array. $a$, $b$, $A$ and $B$ sites are characterized by dark eigenstates that are different linear combinations of the three $J=1$ states. Square plaquettes are characterized by a time-reversal breaking flux, $\phi$, which is staggered throughout the lattice. The lattice harbors a two-site unit cell and is invariant under translation by direct lattice vectors $\vec{g_1}$ and $\vec{g_2}$. {\bf (B)} An optimized band-structure in the reduced Brillouin zone (RBZ) depicting a flatness ratio $f \approx 11.5$. The lowest band carries Chern index $C=-1$. The electric field tilt is $\{\Theta_0, \Phi_0 \}=\{ 0.68, 5.83\} $ and the electromagnetic driving parameters are detailed in the supporting information.}
\label{fig:singleparticle}
\end{figure}

To ensure that our hardcore bosons reside in a topological flat band, we adjust the optical beams that dress the molecules to produce a square lattice with four types of sites, $\{a, b, A, B\}$, as shown in Fig.~\ref{fig:singleparticle}A (see supporting information for details). 
Owing to interference between the dressing lasers, the dark state on  each of the sites is a \emph{different} linear combination of the three $J=1$ states, implying that the hardcore boson, $a_i^{\dagger}$, is site-dependent. Despite the existence of four unique lattice sites, so long as $t_{ij}$ and $V_{ij}$ remain invariant under the direct lattice vectors $\vec{g_1}$ and $\vec{g_2}$ (Fig.~\ref{fig:singleparticle}A), the Hamiltonian retains a two-site unit cell. Thus, computing the single-particle band-structure produces two bands in momentum space, as shown in Fig.~2B. The nonzero Chern number, $C=-1$, of the bottom band reflects the breaking of time-reversal  arising from the asymmetry between the intensity of  left- and right-circularly polarized light fields \cite{Yao12}.

To characterize the single-particle dispersion, we compute the flatness ratio, $f$, between the band-gap and the width of the lowest band \cite{Sun11,Tang11,Neupert11}. 
Numerical optimization of the electric-field and the optical  dressing yields a variety of flat bands with flatness ratio, $f >10$. 
Since an applied DC electric field changes the transition dipole moments between single-molecule eigenstates,  flatness optimization must be re-performed at each field strength (if one desires to retain a flat-band). 
The optimized band-structure depicted in Fig.~2B  has $f \approx 11.5$  and is obtained at weak DC electric fields, just strong enough to split the degeneracy within the $J = 1$ manifold (relative to the dipolar interaction strength) and to set the quantization axis \cite{Chotia12}.  In addition to the experimental simplicity of working with weak DC fields, such a scenario also effectively eliminates direct long-range interactions between the hardcore bosons, as the induced  dipole moments of $\ket{\uparrow}$ and $\ket{\downarrow}$ are then negligible.


\begin{figure*}
\begin{center}
\includegraphics[width=0.6\textwidth]{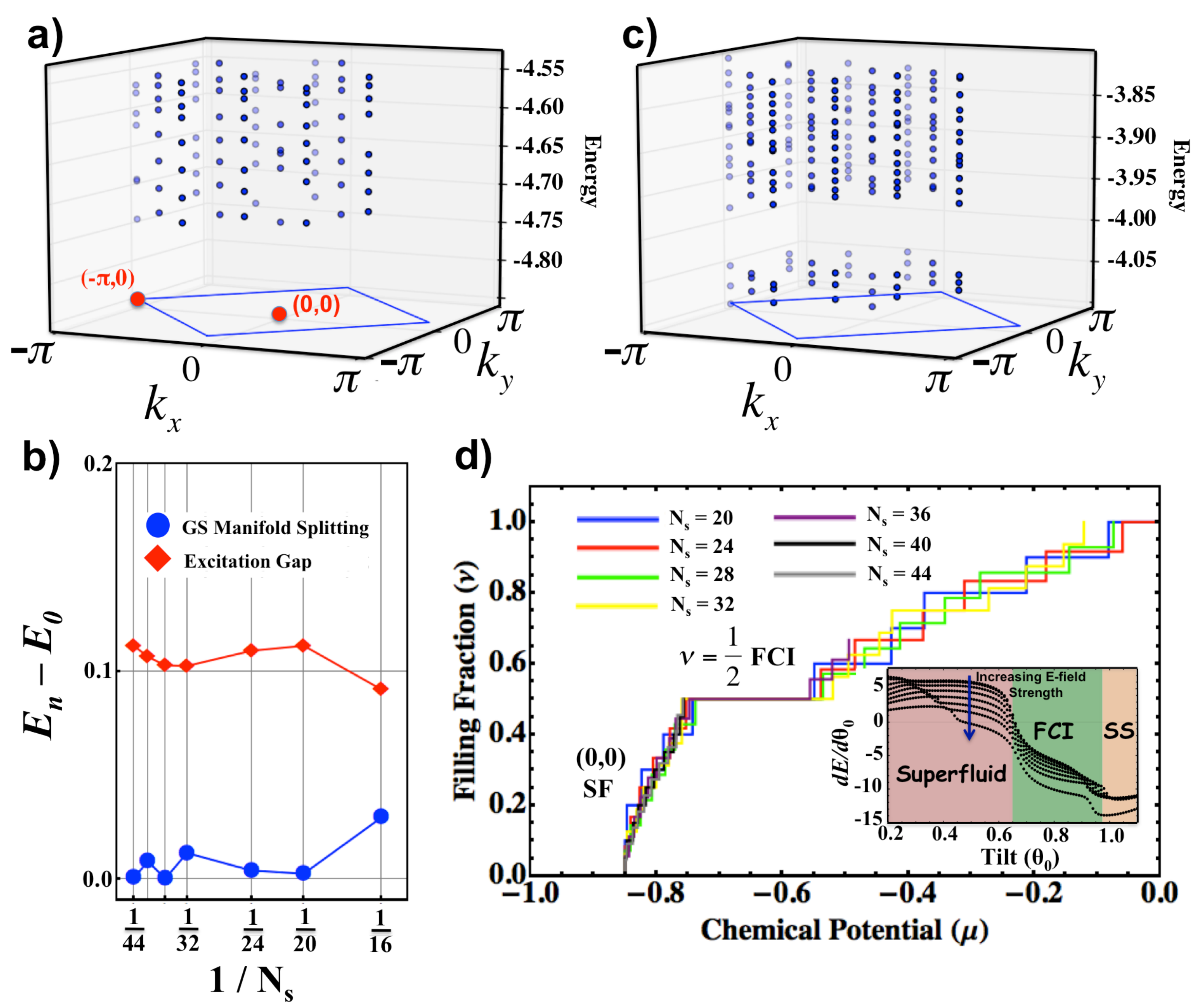}
\end{center}
\caption{\textbf{Evidence for $\nu=1/2$ FCI state.}~{\bf (A)} Exact diagonalization of the full Hamiltonian at $\nu=1/2$ with a total of $N_s=24$ sites and $N_b=6$ hardcore bosons. The electric field and driving parameters are identical to those used in Fig.~2b. To avoid self-interaction, we truncate the dipolar terms at order $1/(3R_0)^3$.  There exist two degenerate ground states in momentum sectors $(k_x,k_y)=(0,0)$ and $(k_x,k_y)=(-\pi,0)$ consistent with a $\nu=1/2$ FCI state on a torus ($k_x,k_y$ are crystal momenta).~{\bf (B)} Finite size scaling from $N_s = 16$ to $N_s = 44$ suggests a stable spectral gap in the thermodynamic limit.~{\bf (C)} Quasihole counting for the same parameters as in (a) with a single boson removed ($N_b=5$). There exists a clear gap below which there are $36$ low-energy quasihole states, consistent with our analytical counting formula  \cite{Regnault11}.~{\bf (D)}  Filling fraction as a function of chemical potential, $\mu = E_{N_b+1}-E_{N_b}$ (where $E_{N_b}$ is the ground state energy with $N_b$ bosons). 
Below $\nu=1/2$, there exists evidence of a clear compressible superfluid state, while at $\nu=1/2$, there exists a plateau indicative of an incompressible quantum liquid. This plateau can also be interpreted as a magnetization plateau in the language of frustrated magnetism \cite{Laughlin87, Kalmeyer89}. (inset) Depicts $dE/d\Theta_0$  as a function of tilt $\Theta_0$. Phase transitions between the superfluid, Chern insulator and striped solid are evidenced as jumps in $dE/d\Theta_0$. Curves from top to bottom are for increasing electric field strength from $E = 0.4-8$ kV/cm.  }
\label{fig:fqhevidence}
\end{figure*}

With topological flat bands in hand, we now consider the actual many-body phases which arise at finite lattice filling fractions $\nu$ (number of spin flips per unit cell). 
To this end, we perform exact diagonalization of the full many-body Hamiltonian at $\nu=1/2$ on systems of varying sizes up to $N_s = 44$ sites with periodic boundary conditions. 
For weak electric fields tilted near the so-called magic angle, $\Theta_0 = \cos^{-1}(1/\sqrt{3})$ where the strong phase anisotropy between $\hat{X}$- and $\hat{Y}$- hops produces particularly flat Chern bands, diagonalization reveals the existence of a bosonic $\nu=1/2$ fractional Chern insulator.
This phase exhibits gapped fractionalized quasiparticles of effective charge $1/2$ with respect to the microscopic spin-flip excitations.
As numerical diagnostics, these topological features require the presence of two-fold ground state degeneracy on a torus (Fig.~\ref{fig:fqhevidence}A) and a neutral spectral gap that is stable as the system size increases (Fig.~\ref{fig:fqhevidence}B). 
The quantity analogous to the Hall conductance, $\sigma_{xy} =   \frac{1}{2\pi} \int \int F(\theta_x, \theta_y) d\theta_x d\theta_y =  -0.5$, appears unambiguously in the response of the system to boundary-condition twists $\{\theta_x, \theta_y\}$ (equivalent to flux insertion) in the form of a well-quantized many-body Berry curvature, $F(\theta_x, \theta_y) = \text{Im} ( \langle \frac{ \partial \Psi}{\partial \theta_y} |  \frac{ \partial \Psi}{\partial \theta_x} \rangle - \langle \frac{ \partial \Psi}{\partial \theta_x} |  \frac{ \partial \Psi}{\partial \theta_y} \rangle)$.

The counting statistics of low energy quasihole states provide a direct diagnostic of the fractionalization of removed particles.
In the continuum, this  counting  can be understood through a generalized Pauli principle, in which there should be no more than one particle in two consecutive Landau-level orbitals; interestingly, the same overall counting holds for the lattice-dominated Chern insulator \cite{Regnault11}. 
Counting the total number of admissible quasihole arrangements on a torus (for a $\nu =1/2$ FCI) yields, $Q_{torus} = \binom{N_{uc}+1-N_b}{N_{uc}+1-2N_b}- \binom{N_{uc}-1-N_b}{N_{uc}+1-2N_b}$, where $N_{uc} = N_s/2$ is the number of lattice unit cells and $N_b$ is the number of hardcore bosons.
%
In our system,  numerically counting the total number of quasihole states matches the above formula. For example, the energy spectrum of the system (depicted in Fig.~3A) with a single boson removed ($N_{uc} = 12$, $N_b = 5$) is shown in Fig.~\ref{fig:fqhevidence}C; from the analytic counting, one obtains $Q_{torus} = 36$, in precise agreement with the number of states below the spectral gap.

Remaining at $\nu = 1/2$, we now probe the many-body phases which arise as one varies the DC  field strength and  the tilt, $\Theta_0$, while adjusting the optical parameters to keep the local dark states fixed (see supporting information for details).
Changes in the tilt alter the geometry of the dipoles and introduce additional dispersion into the single-particle bands. 
On the other hand, increasing the electric field strength enhances the long-range interactions.
These qualitative differences in the microscopics yield a rich phase diagram exhibiting both conventional and topological phases, as shown in Fig.~\ref{fig:mscheme}C. 
In addition to the FCI phase, there exist four distinct crystalline phases at strong DC fields and a large superfluid region at moderate fields. While we use the language of lattice bosons above, we note that the FCI phase may also be interpreted in the language of frustrated magnetism as a chiral spin liquid while the competing superfluid and crystalline phases correspond to XY ordered magnetic and spin density wave phases \cite{Laughlin87, Greiter09}. 
As summarized in Table~I., we characterize the nature of these phases via four diagnostics: i) ground-state degeneracy, ii) spectral flow under magnetic flux insertion (superfluid response), iii) real-space structure factor $\langle n(R) n(0) \rangle$, and iv) $\sigma_{xy}$, the many-body Berry curvature  (see supporting information for details).

\begin{table*}
\centering 
  \caption{Diagnostics of Many-body Phases  } 

\begin{tabular}{c| c| c| c|c}
 \hline\hline Phase & Degeneracy & SF Response & Structure Factor &$\sigma_{xy}$ \\ [0.5ex]	
\hline $\nu=1/2$ Chern insulator&2&none&fluid&-0.5 \\ 
Superfluid&1&isotropic&fluid&gapless \\ 
Striped supersolid (SSS)&3&uni-directional&stripes&gapless \\ 
Knight's Move solid (KMS) & 4 & none &knight's move& 0 \\ 
Checkerboard solid (CKB) & 2 & none &checkerboard& 0 \\ 
Striped Solid (SS) & 4 & none&stripes& 0 \\ [1ex] \hline
 \end{tabular} \label{table:nonlin} 
\end{table*}

Next, we consider a possible route to  preparing the $\nu=1/2$ fractional Chern insulator. 
In current polar molecule experiments, the spin-flip ``vacuum'', corresponding to all sites in the $\ket{\downarrow}$ state, may be prepared with high fidelity from Feshbach molecules by two-photon stimulated
Raman adiabatic passage  \cite{Chotia12,Aikawa10,Deiglmayr08}.  Moreover, rotational excitations of the molecules  can be deterministically introduced by electromagnetically driving the $\ket{\downarrow}$-to-$\ket{\uparrow}$ transition, enabling the straightforward preparation of finite density, low-temperature states within the superfluid and crystalline phases. A detailed variational mean-field study confirms this and generates a non-topological phase diagram consistent with the exact diagonalization (see supporting information). 
The topological order of the FCI state impedes direct preparation by an analogous procedure. 
If the phase boundaries surrounding the FCI state are second order, one might attempt to prepare this state by adiabatically tuning the electric field across the  transition. 
The only known second-order transition between a superfluid and the $\nu=1/2$ FCI phase is multicritical \cite{Barkeshli12}, which suggests that this  phase boundary is generically first order,  consistent with the numerics presented in the inset of Fig.~3D.
On the other hand, continuous Mott insulator to FCI transitions are less finely tuned and may constitute a promising avenue for preparation. 
In particular, the striped solid phase may be reduced to a simple non-translation-symmetry breaking Mott insulator in the presence of a one-dimensional superlattice potential. 
This observation is consistent with the existence of a weaker, finite-size, cross-over at the FCI to striped-solid phase boundary (inset Fig.~3D).




The most direct experimental realization of our proposal would be in molecules with a ${}^1\Sigma$ ground state and no hyperfine structure. With quantum degenerate Bose gases of multiple nuclear-spin-free isotopes of 
Sr and Yb 
 readily available \cite{Daley11}, such hyperfine-free SrYb molecules  seem likely to be realized in the near-future. 
 However, our proposal can also be carried out in currently available ultracold polar molecules, such as  ${}^{40}$K${}^{87}$Rb \cite{ni08}, ${}^7$Li${}^{133}$Cs \cite{Deiglmayr08}, and ${}^{41}$K${}^{87}$Rb \cite{Aikawa10}.  Here, we focus our discussion on ${}^{40}$K${}^{87}$Rb. For the optically excited states $|e_{1}\rangle$ and $|e_{2}\rangle$, we propose to use the $|J',m'\rangle = |2,\pm 2\rangle$ rotational states of the $v' = 41$ vibrational level of the ${(3)}^1\Sigma^+$ electronic state. These states harbor  a strong 640 nm transition to the ground state \cite{Aikawa10,Chotia12}.
To realize our proposal, we require a hierarchy of energy scales corresponding to, $H_{lattice} \lesssim H_{hf} \ll \Omega_i \ll E_{1,0}-E_{1,1}$, where $H_{lattice}$ describes the optical lattice potential and $H_{hf}$ characterizes the molecule's hyperfine structure \cite{Chotia12}.  For ${}^{40}$K${}^{87}$Rb, this hierarchy is easily realized since $H_{hf} \sim 1$ MHz,  while $E_{1,0} - E_{1,1} = 160$ MHz at a moderate DC field strength, $E = B/d \approx 0.5$kV/cm.
By ensuring that the optical dressing ($\Omega_i$) is weak relative to the splitting, $E_{1,0}-E_{1,1}$, we can employ  frequency selection during the creation of the $M$-scheme; meanwhile, the condition $H_{lattice} \lesssim H_{hf} \ll \Omega_i$ allows us  to consider hyperfine and tensor-shift effects   only after the dark state ($\ket{\uparrow}$) is already defined.
Since the composition of the dark state is different on various sites, the spin-flip particle will be subject to  a site-dependent lattice potential and hyperfine structure.  The application of a magnetic field $\sim 10^3$ G (already present in experiments to tune the Feshbach resonance \cite{ni08})  defines the decoupled nuclear spin basis as an approximate eigenbasis of both $\bra{\uparrow}H_{hf}\ket{\uparrow}$ and $\bra{\downarrow}H_{hf}\ket{\downarrow}$. 
It is  then possible to find a pair of nuclear spin eigenstates that have good  overlap and whose compensated energy difference (taking into account $H_{hf}$, $H_{lattice}$, and $t_{ii}$) is  equal on all sites  (see supporting information for details).


As probe light couples directly to the rotational motion of the dipoles,  it is possible to   measure the single spin-flip response of the system in order to detect and then characterize the FCI state. 
For example, the spectral function can be measured at finite energy and momentum using two-photon Bragg spectroscopy, 
 providing direct information regarding fractionalization \cite{Goldman12,Kjall11}. On the edge, one should observe gapless chiral Luttinger liquid behavior, while in the bulk,  the response should exhibit a gap to the multi-quasiparticle continuum. Such a gap manifests as an effective ``magnetization'' plateau as shown in Fig.~3D \cite{Laughlin87}. In this context, the fractionalization of the FCI state would lead to a nontrivial power-law onset of tunneling response. 
Alternatively, time-of-flight absorption imaging can provide  snapshots of the ground state liquid configurations.
Such measurements allow direct exploration of the structure factor of the FCI wavefunction and raise the tantalizing prospect of directly imaging  fractionalization in real space.  
For instance, one can introduce a spin-flip into the bulk, allow the pair of fractionalized quasiparticles to propagate and finally, read-out using full density microscopy.

In summary, we have introduced a scheme by which the collective physics of  dipolar interacting spins can be harnessed to realize  a  fractional Chern insulator. 
While we have focused our discussion on polar molecules, our proposal can be realized in any system composed of electric or magnetic dipolar interacting generalized spins; such degrees of freedom are found in a diverse array of contexts ranging from magnetic atoms and Rydberg ensembles to solid-state spins \cite{Childress06,Lu10}. In particular, for exchange coupled electronic spin dimers or  hyperfine coupled nuclear and electronic spins, one finds an effective level-structure nearly identical to that depicted in Fig.~1B. The dipolar interaction between such coupled spins also yields topologically nontrivial, flat,  spin-flip band-structures, enabling the potential realization of a solid-state Chern insulator. While it is challenging to create electromagnetic gradients on the shortened length scales associated with such solids, it may nevertheless  be possible to implement a flux lattice via  {\emph{homogeneous}} driving, staggered fields and spin-orbit interaction. 
With this in mind, several intriguing directions can be considered, including engineered solid-state defect arrays, magnetic atoms trapped in a quantum solid matrix, and dimer-based magnetic insulators \cite{Childress06, Lu10, Giamarchi08}.
Such approaches promise the advantage of trapping dipoles at closer distances than in an optical lattice, and in the case of quantum solid matrices, such systems are already commonly used for precision laser spectroscopy. 
With the  enhanced interaction strength  in a solid-state environment,  strongly correlated many-body phases could be extremely robust.


We gratefully acknowledge the  insights of P. Zoller, E. Demler and S. Bennett.  We thank A. Chandran, N. Lindner, S. Stellmer, F. Schreck, W. Campbell, A. M. Rey, J. Preskill, K. Hazzard, S. Manmana, E. M. Stoudenmire, S. Todadri and J. Alicea  for helpful discussions.  This work was supported, in part, by the NSF, DOE (FG02-97ER25308), CUA, DARPA, AFOSR MURI, NIST, Lawrence Golub Fellowship, Lee A. DuBridge Foundation, IQIM and the Gordon and Betty Moore Foundation.

\end{document}


\title{Supporting Online Material for

Realizing Fractional Chern Insulators with Dipolar Spins}

\author{N. Y. Yao$^{*}$}
\author{A. V. Gorshkov$^{*}$}
\author{C. R. Laumann$^{*}$}
\author{A. L\"{a}uchli}
\author{J. Ye}
\author{M. D. Lukin}

\maketitle
\begin{center}
$^{*}$ These authors contributed equally to this work.
\end{center}
\section{Deriving the Effective  Hamiltonian}

Here, we derive the effective Hamiltonian, $H_B =  -\sum_{ij} t_{ij} a_i^{\dagger} a_j + \frac{1}{2}\sum_{i \neq j} V_{ij} n_i n_j$ (Eqn.~3 of the main text). The geometry is shown in the inset of Fig.~1A. 
The molecules lie in the $X$-$Y$ plane and the applied DC electric field has spherical coordinates $(\Theta_0,\Phi_0)$ in this basis.  
%
To simplify the notation, we define 
$\ket{\uparrow} = s |1,-1\rangle + v  |1,1\rangle + w  |1,0\rangle$, where $s=\Omega _2 \Omega _4/\tilde{\Omega} $, $v=\Omega _1 \Omega _3/\tilde{\Omega} $, $w=-\Omega _1\Omega _4/\tilde{\Omega} $.
Consider two dressed molecules at positions $i$ and $j$ separated by $\mathbf{R} = (R,\theta, \phi)$ (spherical coordinates in the $\{x,y,z\}$ basis). 
The dipolar interaction between these molecules can be written in spherical tensor form as:
\ba
H_\textrm{dd} &=& -\frac{1}{4\pi\epsilon_0} \frac{\sqrt{6}}{R^3} \sum_{q = -2}^{2} (-1)^q C_{-q}^2(\theta, \phi) T^2_q(\mathbf{d}^{(i)}, \mathbf{d}^{(j)}), \label{eq:dd}
\ea
where $C^k_q(\theta, \phi)$ is the spherical harmonic of degree $k$ and $z$ angular momentum $q$ in the normalization of \cite{Brown03}.
Here,  $T^2$ is the rank $2$ spherical tensor generated from the dipole operators; in particular,  
$T^2_{\pm2}(\mathbf{d}^{(i)}, \mathbf{d}^{(j)}) = d^{(i)}_{\pm}  d^{(j)}_{\pm}$, 
$T^2_{\pm1}(\mathbf{d}^{(i)}, \mathbf{d}^{(j)}) = \left(d^{(i)}_z  d^{(j)}_{\pm} + d^{(i)}_{\pm}  d^{(j)}_z\right)/\sqrt{2}$, 
$T^2_0(\mathbf{d}^{(i)}, \mathbf{d}^{(j)})= \left(d^{(i)}_-  d^{(j)}_{+} + 2 d^{(i)}_z  d^{(j)}_z + d^{(i)}_{+}  d^{(j)}_-\right)/\sqrt{6}$, and $d_{\pm} = \mp(d_x \pm i d_y)/\sqrt{2}$ \cite{Brown03}.  
Expressing the dipolar interaction in this form allows us to   isolate  energy conserving terms. 
We assume that the energy difference between $|1,0\rangle$ and $|1,\pm1\rangle$ ($\Delta$ in Fig.~1B) is larger than the scale of the dipole-dipole interactions. 
Under this assumption, $T^2_{\pm1}$ terms of the dipolar interaction are energy non-conserving and thus highly suppressed.  


We now consider the three resonant contributions to the hopping ($t_{ij}$) matrix elements \cite{Yao12,Gorshkov11, Micheli06,Aldegunde08},
\ba
\bra{ \uparrow_i \downarrow_j } T^2_0 \ket{ \downarrow_i \uparrow_j} &=&\sqrt{\frac{2}{3}} [d_{00}^2 w_i^{*} w_j - \frac{1}{2} d_{01}^2 (v_i^{*} v_j + s_i^{*}s_j)], \\
\bra{ \uparrow_i \downarrow_j } T^2_{+2}\ket{ \downarrow_i \uparrow_j} &=&  -  d_{01}^2 (v_i^{*}s_ j ), \\
\bra{ \uparrow_i \downarrow_j } T^2_{-2}\ket{ \downarrow_i \uparrow_j} &=&  -  d_{01}^2 (s_i^{*}v_ j ), 
\ea
where $d_{00} = \langle 1,0|d_z |0,0\rangle$  and  $d_{01} = \langle 1,\pm1|d_{\pm} |0,0\rangle$. 
Combined with the spherical harmonic coefficients of Eq.~\ref{eq:dd}, these terms determine the directionally dependent hopping $t_{ij}$ of the spin flips. 

The interactions $V_{ij}$ between the spin flips arise as a consequence of the induced  dipole moment which each molecule acquires in an applied electric field. 
Thus, $V_{ij}= \bra{ \uparrow_i\uparrow_j } H_{dd} \ket{ \uparrow_i\uparrow_j} + \bra{ \downarrow_i\downarrow_j } H_{dd}\ket{ \downarrow_i\downarrow_j} -  \bra{ \uparrow_i\downarrow_j } H_{dd} \ket{ \uparrow_i\downarrow_j} - \bra{ \downarrow_i\uparrow_j } H_{dd}\ket{ \downarrow_i\uparrow_j}$ can be calculated in a similar fashion \cite{Yao12}.
First, let us define the induced  dipolar moment of a molecule on site $i$,
$d_{\uparrow_i} = d_1 (|s_i|^2 + |v_i|^2) + \mu_0 |w_i|^2$,
where $d_1=\langle 1,\pm1|d_z |1,\pm1\rangle$ and $\mu_0= \langle 1,0|d_{z} |1,0\rangle$ that of the $\ket{1,0}$ state.  
The contributing terms to $V_{ij}$ are then (suppressing $ij$ superscripts in ${\bf d }$):  
\ba
\bra{ \downarrow_i \downarrow_j } d_z d_z + \frac{1}{2} (d_+ d_- + d_- d_+) \ket{ \downarrow_i \downarrow_j} &=& d_0^2, \\
\bra{ \uparrow_i \downarrow_j } d_z d_z + \frac{1}{2} (d_+ d_- + d_- d_+) \ket{ \uparrow_i \downarrow_j} &=& d_{\uparrow_i}d_0, \\
\bra{ \downarrow_i \uparrow_j } d_z d_z + \frac{1}{2} (d_+ d_- + d_- d_+) \ket{ \downarrow_i \uparrow_j} &=& d_0d_{\uparrow_j}, \\
\bra{ \uparrow_i \uparrow_j } d_z d_z + \frac{1}{2} (d_+ d_- + d_- d_+) \ket{ \uparrow_i \uparrow_j} &=& d_{\uparrow_i} d_{\uparrow_j} -\frac{1}{2} \mu_{01}^2 (s_i w_i^{*}w_js_j^{*} + w_iv_i^{*}v_jw_j^{*}+\text{c.c.}), \\ 
\bra{ \uparrow_i \uparrow_j } d_+d_+\ket{ \uparrow_i \uparrow_j} &=& -\mu_{01}^2 (s_i w_i^{*} w_j v_j^{*} + w_i v_i^{*} s_j w_j^{*}) ,\\
\bra{ \uparrow_i \uparrow_j } d_-d_-\ket{ \uparrow_i \uparrow_j} &=& -\mu_{01}^2 (w_i s_i^{*} v_j w_j^{*} + v_i w_i^{*} w_j s_j^{*}),
\ea
where $d_{0} = \langle 0,0|d_z |0,0\rangle$ and  $\mu_{01}= \langle 1,\pm1|d_{\pm} |1,0\rangle$ is the transition dipole moment between $|1,0\rangle$ and $|1,\pm1\rangle$.
From $H_{dd}$ there also exists an on-site potential $t_{ii} = \sum_{j \neq i} ( \bra{ \downarrow_i\downarrow_j } H_{dd}\ket{ \downarrow_i\downarrow_j}-\bra{ \uparrow_i\downarrow_j } H_{dd} \ket{ \uparrow_i\downarrow_j})$ which varies between sites; however, as we will see below, inhomogeneities in $t_{ii}$ can be regulated using optical lattice tensor shifts. Finally, we note that we have dropped a uniform chemical potential term associated with the molecule's rotational constant $2B$ (see Fig.~1B of the maintext).

To obtain topological flat bands, we adjust the optical radiation to generate four different types of sites $\{a,b,A,B\}$ (in relation to the M-scheme).
By restricting the variation of $\ket{\uparrow}$ on ($a$ vs.~$A$) sites and on ($b$ vs.~$B$) sites, we ensure that both $t_{ij}$ and $V_{ij}$ are  invariant under the direct lattice vectors $\vec{g}_1, \vec{g}_2$, enhancing the symmetry to that of a checkerboard lattice with a  two-site translational unit cell.
This small generalization from a two-site model \cite{Yao12} provides an important minus-sign freedom in the choice of $w$ between lowercase and uppercase letter sites, which we exploit in tuning the Chern band structures.
The freedom can be seen by examining the constraints imposed by requiring $t_{ij}$ and  $V_{ij}$ to be invariant under $a \leftrightarrow A$ and $b \leftrightarrow B$ (\emph{i.e.} translation by $\mathbf{g}_2$). 
In particular, the relevant constraints  allow $w_{a/b} = w_{A/B}$ or $w_{a/b} = - w_{A/B}$ as solutions.

\section{Experimental Implementation in $^{40}$K$^{87}$Rb}

\subsubsection*{Molecular Hyperfine Structure:}

In this section, we consider the complications in our scheme due to the hyperfine structure of diatomic polar molecules such as $^{40}$K$^{87}$Rb \cite{Chotia12,wang10,Ospelkaus10,ni08}. 
The molecular rotational degree of freedom is naturally coupled to the nuclear spins, $I_1=4$ and $I_2=3/2$ of potassium and rubidium. 
The hyperfine Hamiltonian is dominated by the nuclear quadrupole interaction, which has a typical strength $H_{Q} \sim 1$MHz (for $^{40}$K$^{87}$Rb).  This  interaction splits the degeneracy between the $\ket{1 ,\pm1}$ rotational states implying that our workhorse, the $T^2_{\pm2}$ terms of $H_{dd}$, are off-resonant.  
To overcome this issue, one can simply ensure that the optical dressing $\Omega$ (in the $M$-scheme) is much stronger than $H_{hf}$.  This ensures that the hyperfine interaction is unable to couple the dark state  to  other  dressed eigenstates. 

One final issue  to consider  is the particular choice of  nuclear spin states. 
Since the composition of the dark state differs on the four types of lattice sites $\{a,b,A,B\}$, molecules on these sites are subject to slightly different hyperfine potentials; in particular, $\left \langle \uparrow_A \right | H_{hf}  \left | \uparrow_A \right \rangle=\left \langle \uparrow_a \right | H_{hf}  \left | \uparrow_a \right \rangle \neq \left \langle \uparrow_B \right | H_{hf}  \left | \uparrow_B \right \rangle = \left \langle \uparrow_b \right | H_{hf}  \left | \uparrow_b \right \rangle$.
Furthermore, the appropriate nuclear eigenstates will also depend on whether we are considering the rovibrational ground state ($\left | \downarrow \right \rangle$) or the dark state ($\left | \uparrow \right \rangle$); this is because the decoupled nuclear spin basis (so-called Paschen-Bach regime \cite{Aruldhas97})  is only valid in the first case. 
One can solve this issue by applying a static magnetic field $\sim 10^{3}$ G along the direction of the DC electric field, ensuring that the decoupled basis is appropriate for both $\left | \downarrow \right \rangle$ and $\left | \uparrow \right \rangle$ (note that such a field is already present in experiments such as \cite{ni08}).
Then, it only remains to choose a pair of nuclear eigenstates which have reasonable overlap and resonant energies. 
We have numerically verified that this is  generically achievable. 

\subsubsection*{Optical Lattice  Tensor Shifts:}

Similar to the hyperfine potential, $A$-type and $B$-type sites feel different tensor shifts from the optical lattice. 
As alluded to in the main text, these tensor shifts can be exploited to compensate for dipolar induced $t_{ii}$ terms. 
To start, let us consider a single optical field, $E(R,t) = E(R)e^{-i \omega t} + \text{h.c.}$ which we use to create the lattice potential in the $\hat{X}$ direction. 
The optical potential is given by $H_{lattice} = -E(R)^* \alpha(\omega) E(R)$, where $E(R) = |E(R)| \sum_p \beta_p(R) e_p$, $e_p$ is the polarization basis, and $\alpha(\omega)$ is the polarizability tensor of the molecule. 
Recasting the lattice potential in terms of spherical harmonics yields,
\begin{equation}
	\label{eq:quadrupole}
H_{lattice} = -E^2(R) \left [  \frac{2\alpha_{\perp} - \alpha_{\parallel}}{3} + (\alpha_{\parallel}-\alpha_{\perp}) \sum_p C^2_p \gamma_p           \right ]
\end{equation}
where $\alpha_{\parallel}$ is the polarizability along the  internuclear axis, $\alpha_{\perp}$ is the polarizability transverse to the  internuclear axis, $\gamma_0 = |\beta_0|^2-1/3$, $\gamma_{\pm 1} = 1/\sqrt{3} (\beta_0^* \beta_{\pm} - \beta_{\mp}^* \beta_0)$, and $\gamma_{\pm2} = -\sqrt{2/3}\beta_{\mp}^* \beta_{\pm}$. 


In our case, the optical lattice potential seen by $\left | \downarrow \right \rangle$  is,
\begin{equation}
	\label{eq:quadrupole}
\bra{ \downarrow } H_{lattice} \ket{ \downarrow} = -E^2(R) \left [  \frac{2\alpha_{\perp} - \alpha_{\parallel}}{3} + (\alpha_{\parallel}-\alpha_{\perp}) \bra{ 0,0 } C^2_0 \ket{ 0,0} \gamma_0       \right ]
\end{equation}
while the potential seen by  $\left | \uparrow \right \rangle$ is, 
\begin{eqnarray}
\bra{ \uparrow } H_{lattice} \ket{ \uparrow} &=& -E^2(R)  [  \frac{2\alpha_{\perp} - \alpha_{\parallel}}{3} + (\alpha_{\parallel}-\alpha_{\perp})  \{   \gamma_0  (|s|^2 \bra{ 1,-1 } C^2_0 \ket{ 1,-1} \nonumber \\
 &+& |v|^2 \bra{ 1,1} C^2_0\ket{ 1,1} + |w|^2 \bra{ 1,0 }C^2_0 \ket{ 1,0}) + \gamma_{2} sv^* \bra{ 1,1} C^2_2\ket{ 1,-1} \nonumber \\  
 &+&  \gamma_{-2} s^* v \bra{ 1,-1} C^2_{-2}\ket{ 1,1} \}       ].
\end{eqnarray}
The energy difference $\delta E =\bra{ \uparrow } H_{lattice} \ket{ \uparrow} - \bra{ \downarrow } H_{lattice} \ket{ \downarrow}$ varies between $A$-type and $B$-type sites since the dressing parameters $\{s,v,w\}$ are site-dependent.
The goal is to use this tunable tensor shift to compensate for  dipolar induced $t_{ii}$  terms. 
Note that we can achieve propagation of the optical beams along any direction using only $\sigma_+$ and $\pi$ light.  
Since  our optical field never contains any $\sigma_-$ polarization, we find that $\gamma_{\pm 2}$ terms are zero. Moreover, we have also dropped $\gamma_{\pm 1}$ terms, since $\Delta = E_{1,0}-E_{1,1} \gg H_{lattice}$.
Combining the optical fields along the $\hat{x}$, $\hat{y}$ and $\hat{z}$ direction, we have numerically verified that by simply adjusting the intensities of the lattice light, we can fully compensate  for any inhomogeneous on-site potential.

Finally, we demonstrate  a simple configuration of Raman lasers (with wavelength $\lambda_0$), which generates the M-scheme for the $\{a,b,A,B\}$ checkerboard lattice shown in Fig.~2A.
We take the lattice constant to be $\lambda_L = R_0$ (Fig.~2A) and assume that $\lambda_0 \leq \lambda_L$; this can always be accomplished by  increasing $\lambda_L$ (at the expense of weaker dipolar interactions).
We can tilt the $k$-vectors propagating along $\hat X$ and along $\hat Y$ up or down out of the XY plane to give them a periodicity of $\lambda_L$ (in the XY plane). 
Similarly, we can tilt the $k$-vectors propagating along $(\hat X \pm \hat Y)$ up or down out to give them a periodicity of $\sqrt{2} \lambda_L$ (in the XY plane). 
By using only four out of these eight beams and linearly polarizing them along $\hat k \times \hat z$, we can obtain arbitrary $\Omega_2$ and $\Omega_3$ on A and B sites with $\Omega_1=\Omega_4=0$. By utilizing the other four beams, we can obtain arbitrary $\Omega_1=\Omega_4$ on A and B sites (and their negatives on a and b sites) with  $\Omega_2=\Omega_3=0$. This immediately enables us to construct the four-site $M$-scheme. 


\begin{figure}
\hspace*{-0.4in}
\includegraphics[width=1.1\textwidth]{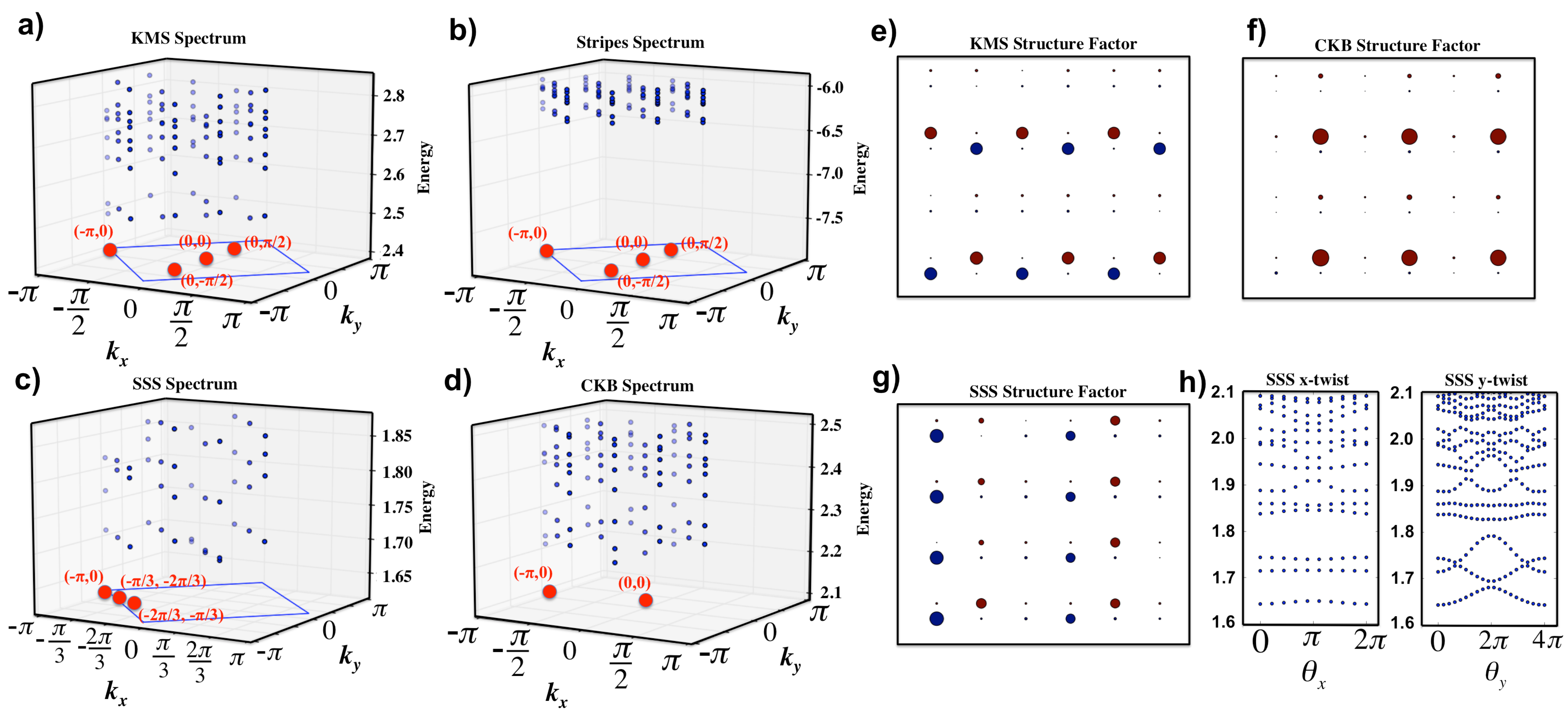}
\caption{\textbf{|- Many-body Phases.}
\textbf{\ (A)} Depicts the spectrum  associated with the knight's move solid phase in the reduced Brillouin zone ($k_x$ and $k_y$ are crystal momenta). The electric field tilt is $\Theta_0 = 0.05$ and the field strength is $|E| \sim 32$kV/cm.   \textbf{(B)} depicts the spectrum  associated with the striped  phase ($\Theta_0 = 1.05$ and $|E| \sim 28$kV/cm).  \textbf{(C)} depicts the spectrum  associated with the striped supersolid phase ($\Theta_0 = 0.68$ and $|E| \sim 36$kV/cm).  \textbf{(D)} depicts the spectrum  associated with the checkerboard phase ($\Theta_0 = 0.25$ and $|E| \sim 40$kV/cm).  \textbf{(E-G)} depicts the structure factor of the KMS, CKB, SSS respectively for the same parameters as above (SS phase omitted).  \textbf{(H)} depicts the spectral flow under magnetic flux insertion of the SSS phase. $\hat{X}$-boundary condition twists produce little dispersion in the ground state energy while $\hat{Y}$ twists yield markedly superfluid behavior. Boundary conditions twists in both $\hat{X}$ and $\hat{Y}$ directions yield minimal superfluid response for  the KMS, CKB and SS phases (not shown).  }
\label{fig:InitializationFidelity}
\end{figure}

\section{Many-Body Phases}

Here, we provide a  detailed description of the many-body phases which arise as one tunes the electric field parameters. First, we note that the field and dressing parameters for the phase diagram (Fig.~1C) are different than those for Figures 2B and 3. This is because the richest many-body phase diagram that we observe does not occur for the band structure with the largest flatness ratio. The band-structure depicted in Fig.~2B occurs at  electric field tilt $\Theta_0=0.68, \Phi_0=5.83$, with optical dressing parameters: $\{\theta_a, \theta_b, \phi_a,\phi_b, \alpha_a,\alpha_b, \gamma_a,\gamma_b\} = \{0.53, 0.97, 1.36, 3.49, 2.84, 2.03, 4.26, 3.84\}$, where we have parametrized: $s_i = \sin( \alpha_i ) \sin (\theta_i) $, $v_i = \sin( \alpha_i ) \cos (\theta_i) e^{i\phi_i} $ and $w_i= \cos( \alpha_i ) e^{i\gamma_i}$.

\begin{figure}
\begin{center}
\includegraphics[width=0.5\textwidth]{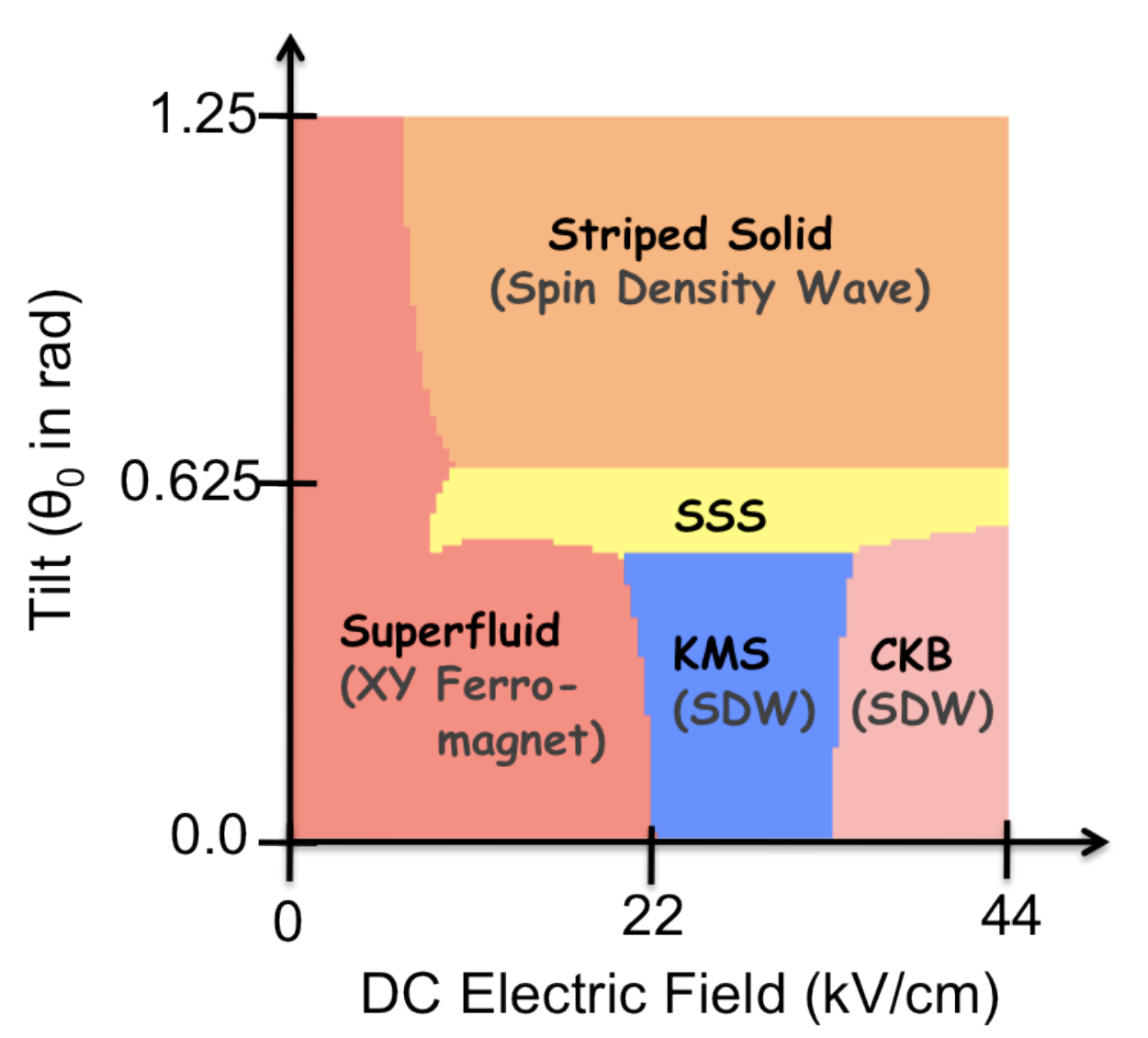}
\end{center}
\caption{\textbf{|- Variational Mean-field theory.} Depicts the variational mean-field theory for ${}^{40}$K${}^{87}$Rb as a function of applied electric field strength and tilt angle $\Theta_0$. We consider twelve variational ansatz's including all allowed solids up to a quadrupled unit cell. The superfluid ansatz is optimized with respect to its winding and a relative phase difference between $A$- and $B$-type sites.  }
\label{fig:MFT}
\end{figure}
The phase diagram shown in Fig.~1C is computed by exact diagonalization for  filling fraction $\nu=1/2$ and a total of $N_s = 24$ sites (parameters: $\{\Theta_0, \Phi_0, \theta_a, \theta_b, \phi_a,\phi_b, \alpha_a,\alpha_b, \gamma_a,\gamma_b\} = \{0.65, 3.68, 2.4, 2.97, 6.06, 4.1, 0.97, 2.74, 3.44, 1.74\}$). The associated band-structure has an optimized flatness ratio, $f\approx7$. The dressing parameters on $A$-type and $B$-type sites are identical to those on $a$ and $b$ sites with the exception that $\gamma_A = \pi + \gamma_a$ and $\gamma_B = \pi + \gamma_b$ (\emph{ie.} $w_{a/b} = - w_{A/B}$). 
At weak electric fields, $E \lesssim 8$ kV/cm, diagonalization reveals the $\nu=1/2$ Fractional Chern Insulator. By changing both the strength and tilt ($\Theta_0$) of the DC field, one can map out a phase diagram containing both conventional and topological phases. 
To isolate the effects of long-range interactions, we ensure that at each tilt, the flatness ratio remains the same for all DC field strengths. This can be achieved by re-optimizing the dressing parameters for each field strength. Alternatively, this corresponds to ensuring that  $d_{00}=d_{01}$ as the field increases; experimentally, one can realize this by dressing the $\ket{1,\pm1}$ states with a long-lived metastable excited state \cite{Chotia12,wang10,Ospelkaus10,ni08}. 
Numerically, we implement this constraint by taking $s_i \rightarrow s_i d_{00}/d_{01}$ and $v_i \rightarrow v_i d_{00}/d_{01}$. 
%

There exist four crystalline phases at strong DC electric fields whose diagnostics we depict in Fig.~S1. At low and intermediate DC field strengths, we observe a large superfluid region. 
This phase is characterized by a unique ground state (typically in either the $(0,0)$ or $(-\pi,0)$ momentum sectors) and a fluid-like real space structure factor. 
The homogeneity of the superfluid changes as one adjusts the DC field strength and tilt. 
At very weak fields $|E| < 4$kV/cm, $A$-type and $B$-type sites are equally populated; however, as one increases the field strength, the anisotropy of the long-range dipolar interaction yields anisotropy in the structure factor. To verify the nature these non-topological phases in the thermodynamic limit, we perform a detailed variational mean-field study. As shown in Fig.~S2, this study confirms not only the existence of these phases, but also the qualitative location of the phase boundaries. Moreover, in the KMS, CKB and SS solids, the mean-field energies match nearly identically with the exact diagonalization energies. This suggests that  product state wavefunctions are valid approximations in these regimes and hence, that one can easily prepare finite density, low-temperature states.